\documentclass[11pt]{article}
\usepackage{epsfig,amsmath, amssymb,a4}

\newenvironment{myeq}[1]{\begin{equation} \hspace{#1}}{\end{equation}}
\newenvironment{mydsp}{\begin{displaymath}}{\end{displaymath}}

\newcommand{\last}[1]{\end{mydsp} \vspace{-0.5cm} \\
                                \begin{myeq}{#1}}
\newcommand{\first}[1]{\end{myeq} \vspace{-0.5cm} \\
                                \begin{mydsp} \hspace{#1}}

\newcommand{\lastl}[2]{\end{mydsp} \vspace{-0.5cm} \\
                                \begin{myeq}{#1} \label{#2}}

\newcommand{\tx}[1]{\textrm{#1}}
\newcommand{\eps}{\varepsilon^{\alpha\beta\mu\nu}}
\newcommand{\no}{\noindent}

\begin{document}

\title{\bf
The $\eta$ $\to$ $\pi^0\pi^0\gamma\gamma$ decay in Generalized
chiral perturbation
theory\setcounter{footnote}{4}\thanks{Presented by M. K. at Int.
Conf. Hadron Structure '02, Herlany, Slovakia, September 22-27,
2002 }}

\author{M.
Koles\'{a}r\setcounter{footnote}{1}\footnote{Marian.Kolesar@mff.cuni.cz}
and J. Novotn\'{y}\footnote{Jiri.Novotny@mff.cuni.cz}
\\ Institute of Particle and Nuclear Physics,\\Charles University,
V Hole\v{s}ovi\v{c}k\'{a}ch 2, 180 00 Czech Republic}
\date{}\maketitle
%   for e-mail address, use macro \email{} in \author - macro:
%   \author{name of the author\email{e-mail address}}{address of the author}

\begin{abstract}{Calculations of $\eta$ $\to$
$\pi^0\pi^0\gamma\gamma$ decay in Generalized chiral perturbation
theory are presented. Tree level and some of next-to-leading
corrections are involved. Sensitivity to violation of the Standard
counting is discussed.}
\end{abstract}

\section{Introduction}

The $\eta(p\,) \to \pi^0(p_1)\,\pi^0(p_2)\,\gamma(k)\,\gamma(k')$ process is a rare decay, which has been recently studied by several authors in context of Standard chiral perturbation theory (S$\chi$PT),
namely at the lowest order by Kn\"ochlein, Scherer and Drechsel \cite{Drechsel} and to next-to-leading by Bellucci and Isidori
\cite{Belucci} and Ametller et al.\cite{Ametller}. The experimental interest for such a process comes from the anticipation of
large number of $\eta$'s to be produced at various facilities.\footnote{according to \cite{Belucci II}, at DA$\Phi$NE about $10^8$
decays per year} The goal of our computations is to add the result for the next-to-leading order in Generalized chiral perturbation theory (G$\chi$PT). The motivation is that one of the important contributions involve the $\eta\,\pi \to \eta\,\pi$ off-shell vertex which is very sensitive to the violation of the Standard scheme and thus this decay provides a possibility of its eventual observation. We have completed the calculations at the tree level, added 1PI one loop corrections and phenomenological corrections of the resonant contribution. These preliminary results we would like to present in this paper.

\section{Kinematics and parameters}

The amplitude of the process can be defined
\begin{equation}
 \langle \pi ^0(p_1)\pi ^0(p_2)\gamma (k,\epsilon )\gamma (k^{^{\prime
}},\epsilon ^{^{\prime }})_{out}|\eta (p)_{in}\rangle =i(2\pi )^4\delta
^{(4)}(P_f-p){\cal M}_{fi}.
\end{equation}
In the square of the amplitude summed over the polarizations
\begin{equation}
\overline{|{\cal M}_{fi}|^2}=\sum_{pol.}|{\cal M}_{fi}|^2,
\end{equation}
 only three independent Lorentz invariant variables occur
\begin{equation}
  s_{\gamma\gamma} = (\:k+k')^2 ,\quad
  s_{\pi\pi} = (p_1+p_2)^2,\quad
  k.P_{\pi\pi} = k.(p_1+p_2),
\end{equation}
where $s_{\gamma\gamma}$ resp. $s_{\pi\pi}$ can be interpreted as
diphoton resp. dipion energy squares. $k.P_{\pi\pi}$ can be
expressed in terms of $s_{\pi\pi},s_{\gamma\gamma}$ and the angle
$\vartheta_{\gamma}$ between the direction of the diphoton and one
of the photons in the center of mass of the diphoton. The range of
the kinematic variables is

%\begin{myeq}{0cm}
%  (k.P_{\pi\pi})_{CMS\; \gamma\gamma}\ =\
%  \frac{1}{4}\,[\,M_{\eta}^2-s_{\gamma\gamma}-s_{\pi\pi}\,] +
%  \frac{1}{4}\,
%  \lambda^{1/2}(M_{\eta}^2,s_{\gamma\gamma},s_{\pi\pi})\,cos\vartheta_{\gamma}.
%\end{myeq}
\begin{equation}
  0 <\, s_{\gamma\gamma} \leq\, (M_{\eta}-2M_{\pi})^2,\quad
  4M_{\pi}^2 \leq\, s_{\pi\pi} \leq\, (M_{\eta}-\sqrt{s_{\gamma\gamma}})^2,\quad
  -1 \leq\, \cos\vartheta_{\gamma} \leq 1\,.
\end{equation}
%\begin{myeq}{0cm}
%  0\ < s_{\gamma\gamma}\ \leq\ (\,M_{\eta}-2M_{\pi}\,)^2
%\more{-0.5cm}
%  4M_{\pi}^2\ \leq\ s_{\pi\pi}\ \leq\ (\,M_{\eta}-\sqrt{s_{\gamma\gamma}}\,)^2
%\more{-2cm}
%  -1\ \leq\ cos\vartheta_{\gamma}\ \leq\ 1
%\end{myeq}
Our goal is to calculate the partial decay width of the $\eta$
particle as the function of the diphoton energy square
$s_{\gamma\gamma}$
\begin{equation}
  \frac{\tx{d}\Gamma}{\tx{d}s_{\gamma\gamma}}\ =\ \frac{1}{(8\pi)^5 M_{\eta}^3}\,
  \int \overline{\rule{0cm}{0.4cm}|{\mathcal{M}}_{fi}|^2}\,
  \lambda^{1/2}(M_{\eta}^2,s_{\gamma\gamma},s_{\pi\pi})
  \sqrt{1-\frac{4M_{\pi}^2}{s_{\pi\pi}}}\;
  \tx{d}\cos\vartheta_{\gamma}\,\tx{d}s_{\pi\pi}.
\end{equation}
%\begin{myeq}{-0.3cm}
%  \tx{d}\Gamma\ =\ \frac{1}{(8\pi)^5 M_{\eta}^3}\,\overline{\rule{0cm}{0.4cm}|{\mathcal{M}}_{fi}|^2}\,
%  \lambda^{1/2}(M_{\eta}^2,s_{\gamma\gamma},s_{\pi\pi})
%  \sqrt{1-\frac{4M_{\pi}^2}{s_{\pi\pi}}}\;
%  \tx{d}cos\vartheta_{\gamma}\,\tx{d}s_{\pi\pi}\,\tx{d}s_{\gamma\gamma}
%\end{myeq}
%\vspace{-0.25cm}
At the lowest order, the S$\chi$PT does not depend on any unknown free order parameters. In contrast, there are two free parameters controlling the violation of the Standard picture in the Generalized scheme. We have chosen them as
%\vspace{-0.25cm}
\begin{equation}
  r\ =\ \frac{m_s}{\hat{m}}\,,\quad X_{GOR}\ =\ \frac{2B\hat{m}}{M_{\pi}^2}
\end{equation}
%\vspace{-0.25cm}
\no and their ranges are $r \sim r_1 - r_2 \sim 6 - 26\, ,\ 0\
\leq\ X_{GOR}\ \leq\ 1$.
\no The Standard values of these
parameters are $r=r_2$ and $X_{GOR}=1$. In the Standard counting
also $\Delta_{GMO}=0$, where
\begin{equation}
  \Delta_{GMO} = \frac{3M_{\eta}^2+M_{\pi}^2-4M_K^2}{M_{\pi}^2}.
\end{equation}
We use abbreviations for
%\begin{myeq}{0cm}
%  r \sim r_1 - r_2 \sim 6 - 26\, ,\quad
%  0\ \leq\ X_{GOR}\ \leq\ 1.
%\end{myeq}
%\no We use abbreviations for
%\vspace{-0.25cm}
\begin{eqnarray}
  \hat{m}\ =\ \frac{m_u+m_d}{2}\,,\ R\ =\
  \frac{m_s-\hat{m}}{m_d-m_u}\\
  r_1 = 2\,\frac{M_K}{M_{\pi}}-1,\ \, r_2 = 2\,\frac{M_K^2}{M_{\pi}^2}-1
 \end{eqnarray}
and
\begin{equation}
\varepsilon = 2\,\frac{r_2-r}{r^2-1}\;.
\end{equation}
%\begin{myeq}{0cm}
%  \zeta\ =\ \frac{Z_0^S}{A_0},\quad r\ =\ \frac{2m_s}{m_u + m_d}.
%\end{myeq}
%\no The range of $r$ is
%\begin{myeq}{0cm}
%  r \sim r_1 - r_2 \sim 6 - 26;\quad
%  r_1 = 2\,\frac{M_K}{M_{\pi}}-1,\ \, r_2 = 2\,\frac{M_K^2}{M_{\pi}^2}-1 .
%\end{myeq}
%\no Although $\zeta$ is a Zweig rule violating parameter and thus it is assumed to be small, its range is unknown. That's why we in our calculation prefer another parametrization with $\zeta$ replaced by a new parameter $X_{GOR}$
%\begin{myeq}{0cm}
%   X_{GOR}\ =\ \frac{2B\hat{m}}{M_{\pi}^2}\ =\ [\,1-\varepsilon-\varepsilon\zeta\,(\,r+2\,)],\quad
%   \varepsilon = 2\,\frac{r_2-r}{r^2-1}
%\more{0cm}
%  0\ \leq\ X_{GOR}\ \leq\ 1.
%\end{myeq}
%\vspace{-0.25cm}

\section{Tree level}

At the $O(p^4)$ tree level, the amplitude has two contributions,
with a pion and an eta propagator. The first one is resonant and
thus we call it `$\pi^0$-pole', the other is nonresonant
`$\eta$-tail'.
%\newpage
%\begin{figure}[h]
%\epsfig{figure=picture.eps,width=0.8\textwidth}
%\end{figure}
The amplitude of the $\pi^0$-pole contribution is
%\begin{myeql}{-2cm}{treepion}
%  i{\mathcal{M}}_{fi}^{(R)}\ =\ \frac{ie^2 K_R}{4\sqrt{3}\pi^2 F_{\pi}^{3}}\; \eps\;
%            \varepsilon^*_{\alpha}(\,k,\lambda)\,
%            \varepsilon^*_{\beta}(\,k',\lambda')\; k_{\mu}k'_{\nu}
%            \; \frac{1}{s_{\gamma\gamma}-M_{\pi}^2}\;,
%\end{myeql}
%\no where ($\Delta_{\pi\eta}=M_{\eta}^2-M_{\pi}^2$)
%\vspace{-0.1cm}
%\begin{myeqn}{-3cm}
%   K_R\ =\ -\,\frac{3\Delta_{\pi\eta}}{4R}\;\alpha_{3\pi\eta}\ =\
%           -\,\frac{3\Delta_{\pi\eta}}{4R}\; \bigg[\
%                  \frac{4\,\delta}{\sqrt{3}}\ +\ \frac{2M_{\pi}^2(r-1)}{\Delta_{\pi\eta}}\;.
%\last{2.5cm}
%      .\;\bigg[\;\frac{2\,(1-X_{GOR})+r\,\varepsilon)}{2+r}\;\Big(1-\frac{4\,\delta}{\sqrt{3}}\;\Big)\
%     +\ \frac{\Delta_{GMO}}{(r-1)^2}\;\Big(1+\frac{4\,\delta}{\sqrt{3}}\;\Big)\,\bigg]\bigg].
%\end{myeqn}
\begin{equation}
\label{treepion}
  i{\mathcal{M}}_{fi}^{(R)}\ =\ -\,\frac{3ie^2\Delta_{\pi\eta}\alpha_R}{16\sqrt{3}\pi^2 F_{\pi}^{3}R}\; \eps\;
            \varepsilon^*_{\alpha}(\,k,\lambda)\,
            \varepsilon^*_{\beta}(\,k',\lambda')\; k_{\mu}k'_{\nu}
            \; \frac{1}{s_{\gamma\gamma}-M_{\pi}^2}\;,
\end{equation}
where $\ \Delta_{\pi\eta}=M_{\eta}^2-M_{\pi}^2\ $,
\begin{equation}
   \alpha_{R}\ =\ \frac{4\,\delta}{\sqrt{3}}\ +\ \frac{2M_{\pi}^2(r-1)}{\Delta_{\pi\eta}}
      \;\bigg[\;\frac{1}{3}\;(\alpha_{\pi\pi}-1)\;\Big(1-\frac{4\,\delta}{\sqrt{3}}\;\Big)\ +\
      \frac{\Delta_{GMO}}{(r-1)^2}\;\Big(1+\frac{4\,\delta}{\sqrt{3}}\;\Big)\,\bigg].
\end{equation}
%\begin{myeqn}{-2cm}
%  K_R\ =\ \frac{M_{\pi}^2}{2R(1-r)(2+r)}\,( (r-1)^2(8+r+5r\varepsilon+r^2\varepsilon-6X_{GOR})\ +
%\last{2cm}
%        +\ (8+2r-r^2)\Delta_{GMO}
%        - 4\sqrt{3}\,\delta( (r-1)^2(2+r\varepsilon-2X_{GOR}) - (2+r)\Delta_{GMO} ))
%\end{myeqn}
%\begin{myeqn}{-1cm}
%  K_R\ =\ \frac{M_{\pi}^2}{2R(1-r)}\, (\,1 + 4\Delta_{GMO} + 3\varepsilon + r^3\varepsilon +
%          6\epsilon\zeta + r^2(\,1+\varepsilon+6\varepsilon\zeta\,)
%\last{2cm}
%          -r(\,2+\Delta_{GMO}+5\varepsilon+
%          12\varepsilon\zeta\,) - 4\sqrt{3}\,\delta(-\Delta_{GMO} + (r-1)^2\varepsilon(1+2\zeta)))
%\end{myeqn}
and $\ \alpha_{\pi\pi}\;=\;1+(6(1-X_{GOR})+3r\varepsilon)/(r+2)\
.$ In the Standard limit $\ \alpha_R,\alpha_{\pi\pi}=1$. $\
\delta/R\ $ is the $\eta-\pi^0$ mixing angle
\begin{equation}
  \delta\ =\ \frac{\sqrt{3}}{4}\;\Big[\,1 + \frac{M_{\pi}^2}{3\Delta_{\eta\pi}}
             \;\big((r-1)^2\,\varepsilon-3\Delta_{GMO}\big)\Big].
%  \delta\ =\ \frac{M_{\pi}^2}{2\sqrt{3}\,(M_{\eta}^2-M_{\pi}^2)}\,
%             (-1+r-\Delta_{GMO}-r\varepsilon+r^2\varepsilon).
\end{equation}
 The structure of the $\eta$-tail amplitude is similar
\begin{equation}
  i{\mathcal{M}}_{fi}^{(\eta)}\ =\ \frac{ie^2 M_{\pi}^2\,\alpha_{\pi\eta}}{12\sqrt{3}\pi^2 F_{\pi}^{3}}\; \eps\;
            \varepsilon^*_{\alpha}(\,k,\lambda)\,
            \varepsilon^*_{\beta}(\,k',\lambda')\; k_{\mu}k'_{\nu}
            \; \frac{1}{s_{\gamma\gamma}-M_{\eta}^2}\; .
\end{equation}
The constant $\alpha_{\pi\eta}$ is
\begin{equation}
  \alpha_{\pi\eta}\ =\ 1\ +\ \frac{(1+2r)(2(1  - X_{GOR})+r\,\varepsilon)}{2+r}\ -\ \frac{2{\Delta }_{GMO}}{r-1}\;.
%  K_{\eta}\ =\ M_{\pi}^2 \,\frac{(r-1)(4 + 2r^2\varepsilon + r(5+\varepsilon)) +
%               (2+2r-4r^2)X_{GOR} - 2(2+r)\Delta_{GMO}}{r^2+r-2}
%  K_{\eta}\ =\ -\frac{M_{\pi}^2}{r-1}\, (\,1 + 2\Delta_{GMO} + \varepsilon + 2\varepsilon\zeta +
%              r\,(-1 + \varepsilon + 2\varepsilon\zeta\,)
%              - 2r^2\,(\,\varepsilon + 2\varepsilon\zeta\,))
\end{equation}

\begin{figure}[h]
\epsfig{figure=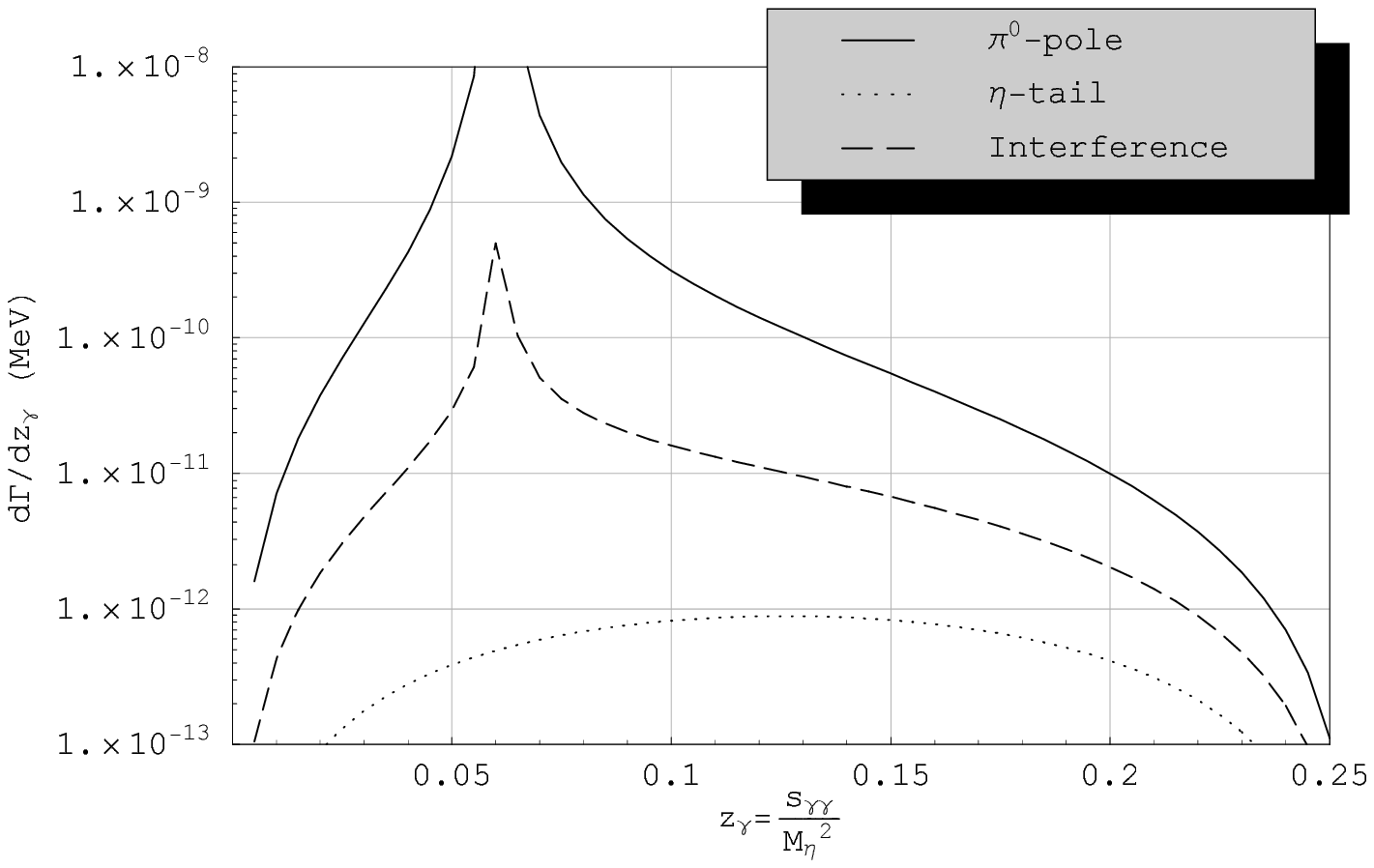,width=0.5\textwidth}
\hspace{-0.2cm}
\epsfig{figure=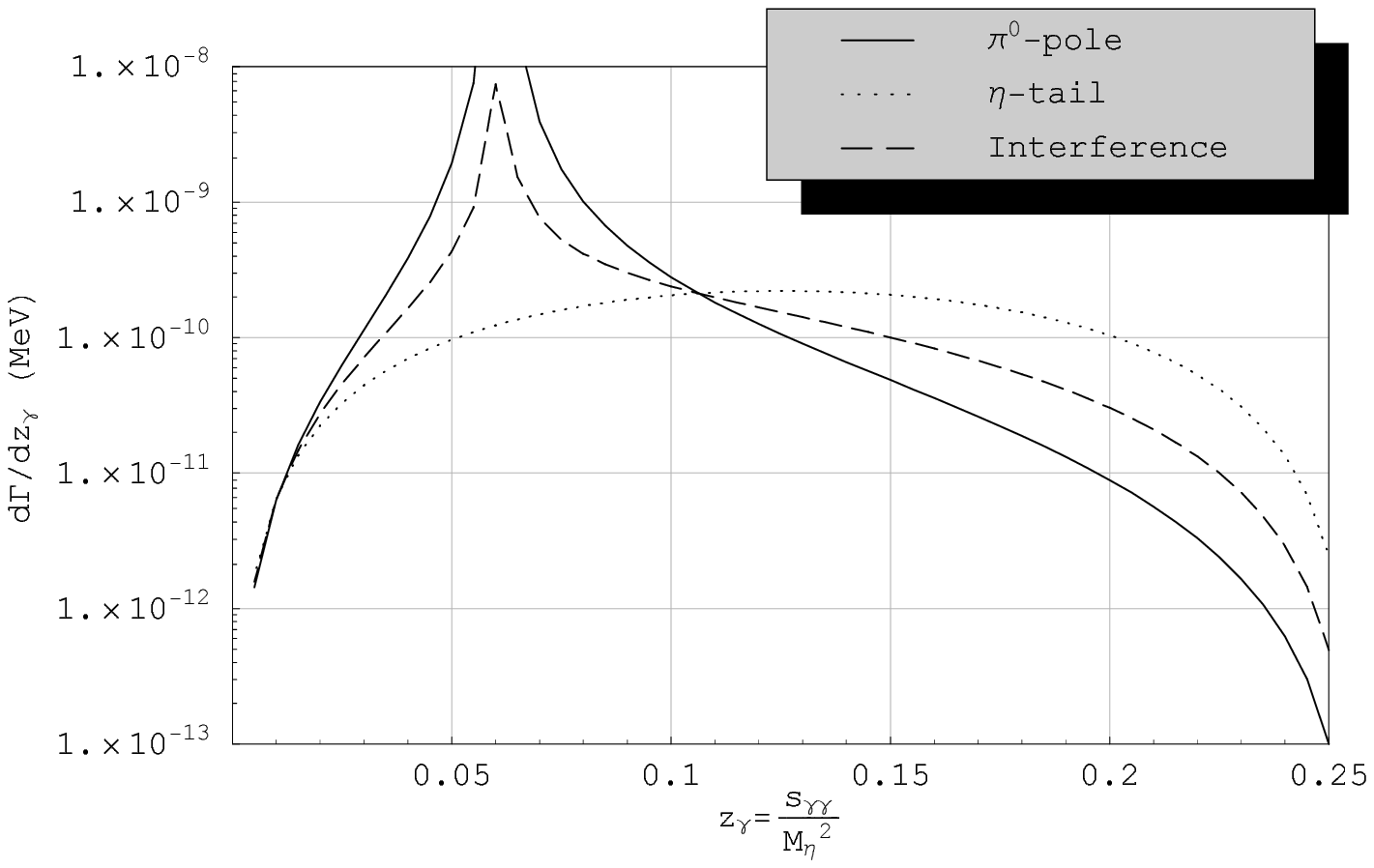,width=0.5\textwidth}
\vspace{-0.5cm}
\caption{S$\chi$PT and G$\chi$PT contributions to the partial decay rate $\tx{d}\Gamma/\tx{d}z_{\gamma}$}
\label{graph4}
\end{figure}

The Standard values of the contributions to the partial decay rate and the maximum possible violation of the Standard counting ($r=r_1,X_{GOR}=0$) are represented in fig.\ref{graph4}. The pole of the resonant contribution at $s_{\gamma\gamma}=M_{\pi}^2\,\sim\,0.06M_{\eta}^2$ is transparent. While in the Standard case it is fully dominant, in the Generalized scheme the $\eta$-tail could be determining in the whole area $s_{\gamma\gamma}>0.11M_{\eta}^2$. The reason can be found in the constant $\alpha_{\pi\eta}$ from the $\eta\,\pi \to \eta\,\pi$ vertex. Its Standard value is equal to $\alpha_{\pi\eta}=1$, but in the Generalized counting it could jump up to $\alpha_{\pi\eta} \sim 16$. Compared to $\alpha_{\pi\eta}$, $\alpha_R$ is not so sensitive, maximum value is $\alpha_R \sim 1.5$ for $r=15,X_{GOR}=1$.

\begin{figure}[h]
\hspace{1.5cm}
\epsfig{figure=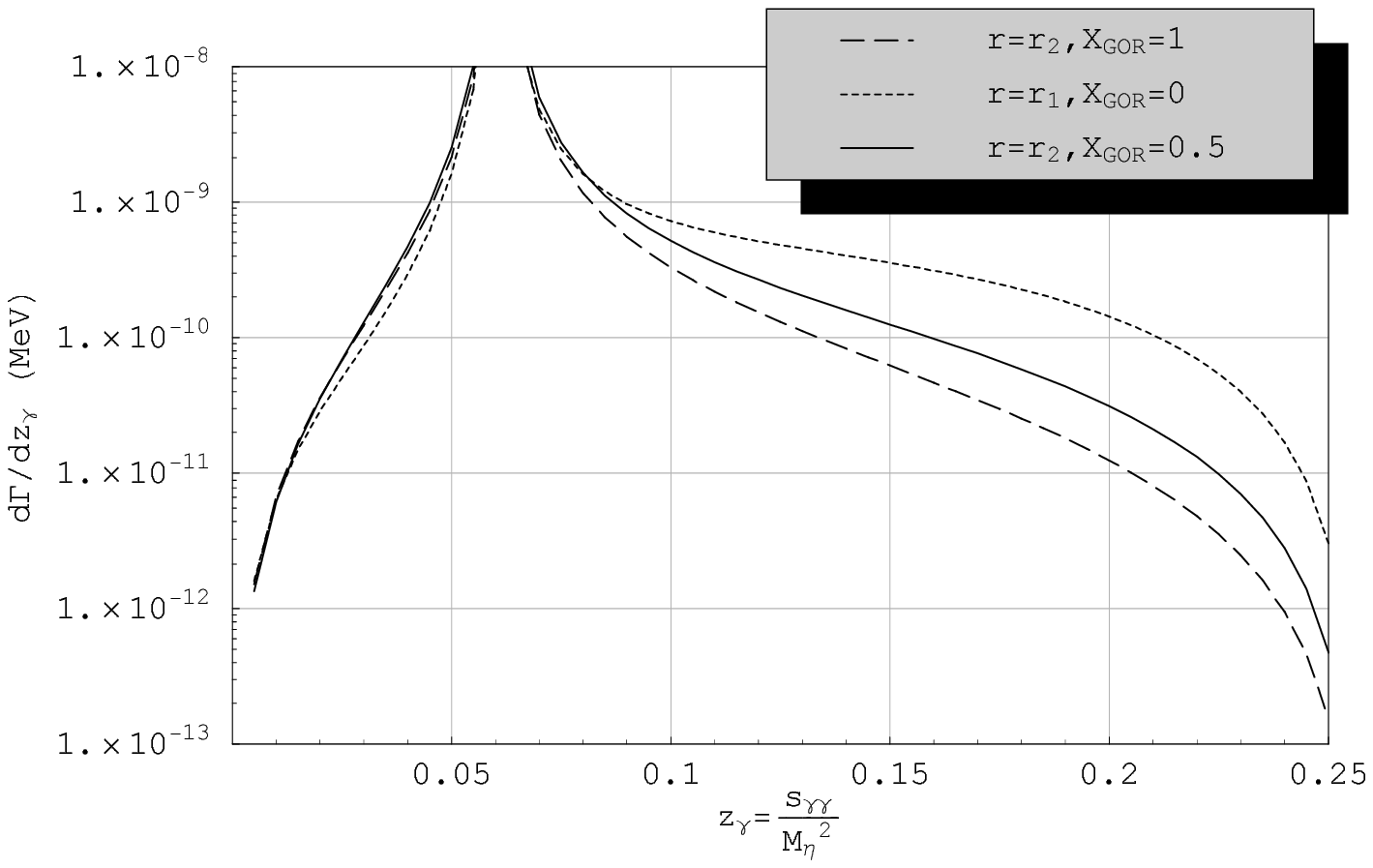,width=0.7\textwidth}
\vspace{-0.25cm}
\caption{Full tree level decay width depending on the parameters $r$ and $X_{GOR}$}
\label{graph6}
\end{figure}

The full decay width for the Standard ($r=r_2,X_{GOR}=1,\Delta_{GMO}=0$) and Generalized case ($r=r_2$,$X_{GOR}=0.5$ and $r=r_1$,$X_{GOR}=0$) is displayed in fig.\ref{graph6}. It can be seen, that even in the conservative intermediate case the change is quite interesting.

\section{One loop corrections}

There are four distinct contributions at the next-to-leading
order: one loop corrections to the $\pi^o$-pole and the
$\eta$-tail, one particle irreducible diagrams and counterterms.
In the latter case we will rely upon the results of
\cite{Ametller}. Their estimate from vector meson dominated
counterterms indicates, that it causes only a slight decrease of
the full decay width. Because the estimate is the same for both
schemes, for our purpose of studying the differences between them
we can leave it for later investigation.

More important might be the corrections to the $\eta$-tail
diagram. At the present time, although we do have the amplitude,
we are still working at the numerical analysis. We decided,
similarly to \cite{Belucci}, to correct the $\pi^0$-pole
amplitude (\ref{treepion}) by a phenomenological parametrization
of the $\eta \to 3\pi^0$ vertex and fix the parameters from
experimental $\eta \to 3\pi^0$ data:
\begin{eqnarray}
  iA_{corr}(\eta \to 3\pi^0)\ =\
  -\;\frac{3i\Delta_{\pi\eta}}{4\sqrt{3}F_{\pi}^{2}R}\,\varrho\, e^{i\phi_R}
  i{\mathcal{M}}_{fi}^{(R)}\;\;\;\;\;\;\;\;\;\;\;\;\;\;\;\;\;\;\;\;\;\;\;\;\;\;\;\;\;\;\;\;\;\;\;\;\\\ =\ -\,\frac{3ie^2\Delta_{\pi\eta}}{16\sqrt{3}\pi^2 F_{\pi}^{3}R}\;
            \varrho\,e^{i\phi_R}\; \eps\;
            \varepsilon^*_{\alpha}(\,k,\lambda)\,
            \varepsilon^*_{\beta}(\,k',\lambda')\; k_{\mu}k'_{\nu}
            \; \frac{1}{s_{\gamma\gamma}-M_{\pi}^2}\;.
\end{eqnarray}
The factor $\varrho$ can be fixed from the experimental results
for $\eta \to 3\pi^0$ \cite{Belucci}
\begin{equation}
  -\,\frac{3\Delta_{\pi\eta}}{4R}\ =\
  M_{K^{\pm}}^2-M_{K^0}^2+M_{\pi^0}^2-M_{\pi^{\pm}}^2,\quad
  \varrho\ =\ 2.0\ \pm\ 0.1\,.
\end{equation}
As the phase $\phi_R$ cancels in the square of the amplitude, it
cannot be determined that way. It is possible to make an estimate
by expanding the $\eta \to 3\pi^0$ one loop amplitude around the
center of the Dalitz Plot \cite{Belucci}. By performing this
procedure in the Generalized scheme it can be found
\begin{equation}
  \phi_R\ =\ \frac{1}{96\pi F_{\pi}^2}\,
  \sqrt{1-\frac{12M_{\pi}^2}{M_{\eta}^2-3M_{\pi}^2}}\
  \Big[\,2M_{\eta}^2+M_{\pi}^2\,\big( 11\,\alpha_{\pi\pi}-2\big)\Big].
\end{equation}
%\begin{myeq}{0cm}
%  \alpha_R\ =\ \frac{1}{96\pi F_{\pi}^2}\,
%  \sqrt{1-\frac{12M_{\pi}^2}{M_{\eta}^2-3M_{\pi}^2}}\
%  [\,2M_{\eta}^2+M_{\pi}^2( 11\,(1 + 3\varepsilon + 6\varepsilon\zeta)-2)]
%\end{myeq}
The 1PI amplitude, where we neglected the suppressed kaon loops,
can be expressed as
%(3s_{\pi\pi}+M_{\pi}^2(\,\frac{8+r+3r\varepsilon-6X_{GOR}}{2+r}-4\,))
\begin{eqnarray}
  i{\mathcal{M}}_{fi}^{(1PI)}\ =\ \frac{e^2}{12\sqrt{3}\pi^2 F_{\pi}^{5}}\; \eps\;
            \varepsilon^*_{\alpha}(\,k,\lambda)\,\varepsilon^*_{\varrho}(\,k',\lambda')\;
            k_{\beta}p_{\nu}\,[\,k'_{\mu}q^{\varrho}-\delta_{\mu}^{\varrho}\;q.k'\,]\nonumber\\
  \times\ [\,3s_{\pi\pi}+M_{\pi}^2(\,\alpha_{\pi\pi}-4\,)]
            \;F(\,k',\,q\,)\ +\ (\,k \leftrightarrow k'),
\end{eqnarray}
%\begin{myeqn}{-4cm}
%  i{\mathcal{M}}_{fi}^{(1PI)}\ =\ \frac{e^2 K}{12\sqrt{3}\pi^2 F_{\pi}^{5}}\; \eps\;
%            \varepsilon^*_{\alpha}(\,k,\lambda)\,\varepsilon^*_{\varrho}(\,k',\lambda')\;
%            k_{\beta}p_{\nu}\ .
%\last{5cm}
%  .\ [\,\delta_{\mu}^{\varrho}\;q.k'-k'_{\mu}q^{\varrho}\,]
%            \;F(\,k',-q\,)\ +\ (\,k \leftrightarrow k')
%\end{myeqn}
where $\,q=p-k\ $ and \vspace{-0.25cm}
\begin{equation}
  F(\,k',\,q\,)\Big|_{{k'}^{2}=0} = \frac{1}{2(q.k')^{2}}\,
  \bigg[\Big(\,\frac{i}{8\pi^{2}}-4M_{\pi}^2\,C_0(k',\,q)\Big)\,q.k' +
  \Big(\,B_0(k',\,q)-B_0(q)\Big)\,q^{2} \bigg].
  \end{equation}
$C_0(\,x,y\,)$ and $B_0(x)$ are the standard scalar two and three
point functions \cite{Veltman}. In the square of the amplitude we
can distinguish the following contributions
\begin{equation}
  \overline{\rule{0cm}{0.4cm}|{\mathcal{M}}_{fi}|^2}\ =\ \sum_{pol.}\,\Big[\
     |{\mathcal{M}}^{(R)}_{fi}|^2 + |{\mathcal{M}}^{(1PI)}_{fi}|^2 +
     {\mathcal{M}}_{fi}^{(1PI-R)} + {\mathcal{M}}_{fi}^{(\eta)\;2}\,\Big]\,,
\end{equation}
where ${\mathcal{M}}_{fi}^{(1PI-R)}$ is the 1PI-resonant
interference and
\begin{equation}
  {\mathcal{M}}_{fi}^{(\eta)\;2}\ =\ |{\mathcal{M}}^{(\eta)}_{fi}|^2 +
  {\mathcal{M}}_{fi}^{(1PI-\eta)} + {\mathcal{M}}_{fi}^{(R-\eta)}
\end{equation}
is the effect of the $\eta$-tail diagram.
${\mathcal{M}}_{fi}^{(1PI-\eta)}$ and
${\mathcal{M}}_{fi}^{(R-\eta)}$ are the interference terms denoted
in the same way as ${\mathcal{M}}_{fi}^{(1PI-R)}$.

\begin{figure}[h]
\epsfig{figure=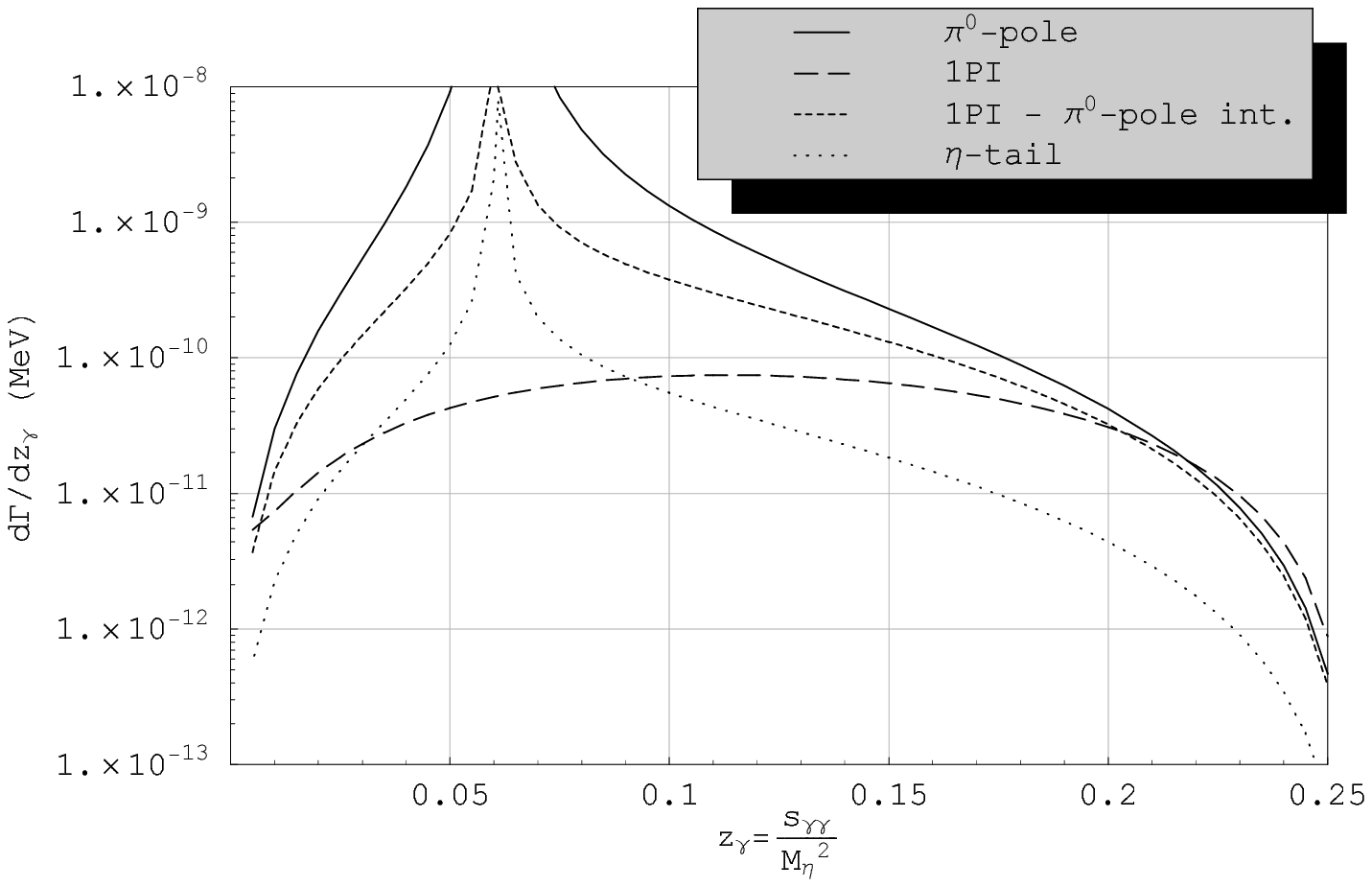,width=0.5\textwidth}
\hspace{-0.2cm}
\epsfig{figure=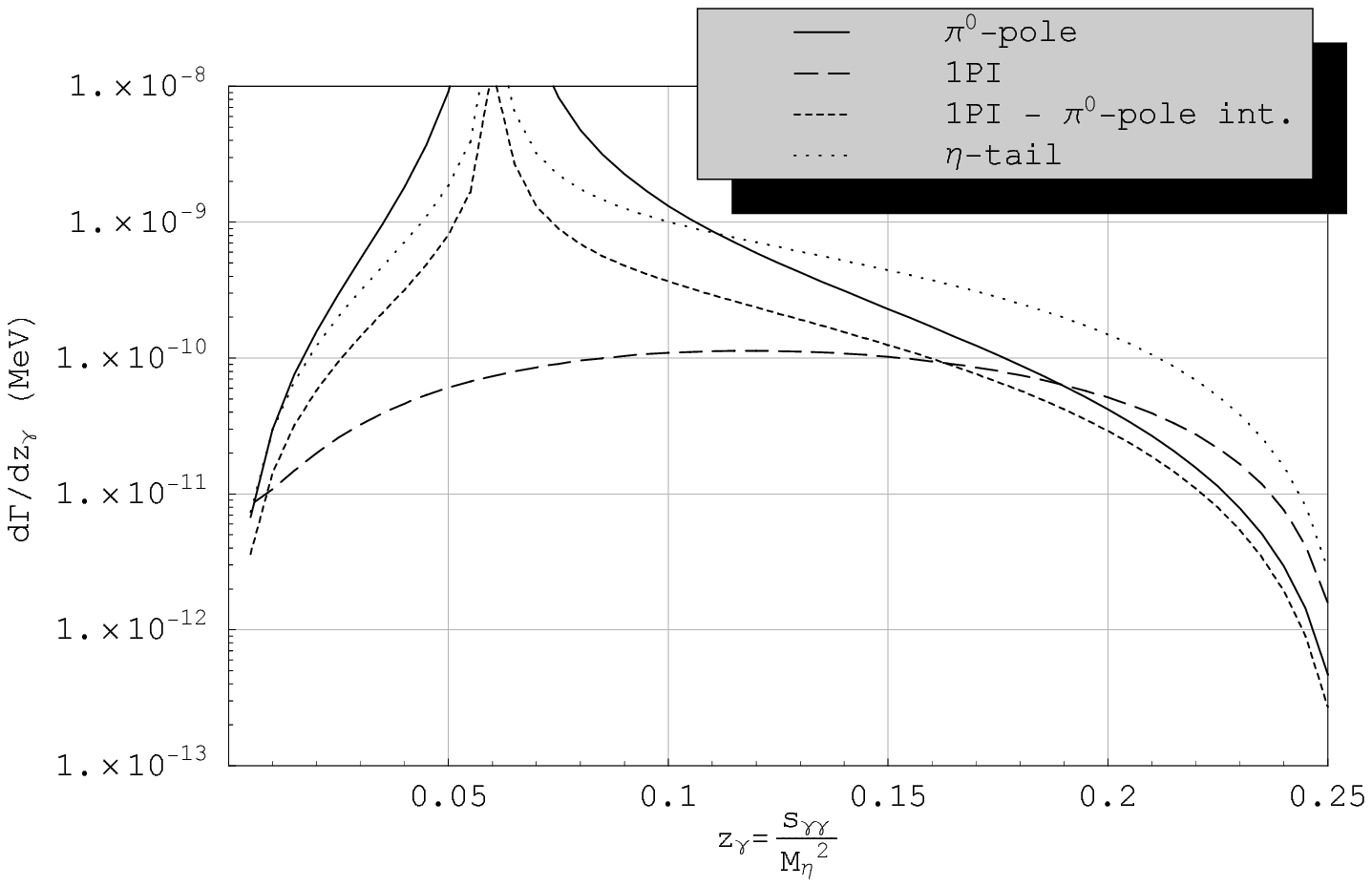,width=0.5\textwidth}
\vspace{-0.5cm}
\caption{S$\chi$PT and G$\chi$PT contributions to the partial decay rate $\tx{d}\Gamma/\tx{d}z_{\gamma}$}
\label{graph1}
\end{figure}

Similarly to the tree level case, fig.\ref{graph1} represents the relevant contributions to the decay width for the Standard and the maximum violation of the Standard scheme. The absolute value of the interference terms is drawn.
Indeed, at the left picture we reproduce the results of
\cite{Drechsel}, \cite{Belucci} and \cite{Ametller}. The
1PI--resonant interference is destructive for
$s_{\gamma\gamma}>M_{\pi}^2$ and constructive otherwise. This
confirms the results of Ametller et al.\cite{Ametller} and is in
contradiction to work of Bellucci and Isidori \cite{Belucci}.
Moving to the Generalized counting, the 1PI graphs does not
dramatically influence the indications of the tree level. Fig.4
displays the corrected decay width in both extremes of the
parameters. For comparison, also the change in 1PI diagrams is
shown.
\begin{figure}[h]
\hspace{1.5cm}
\epsfig{figure=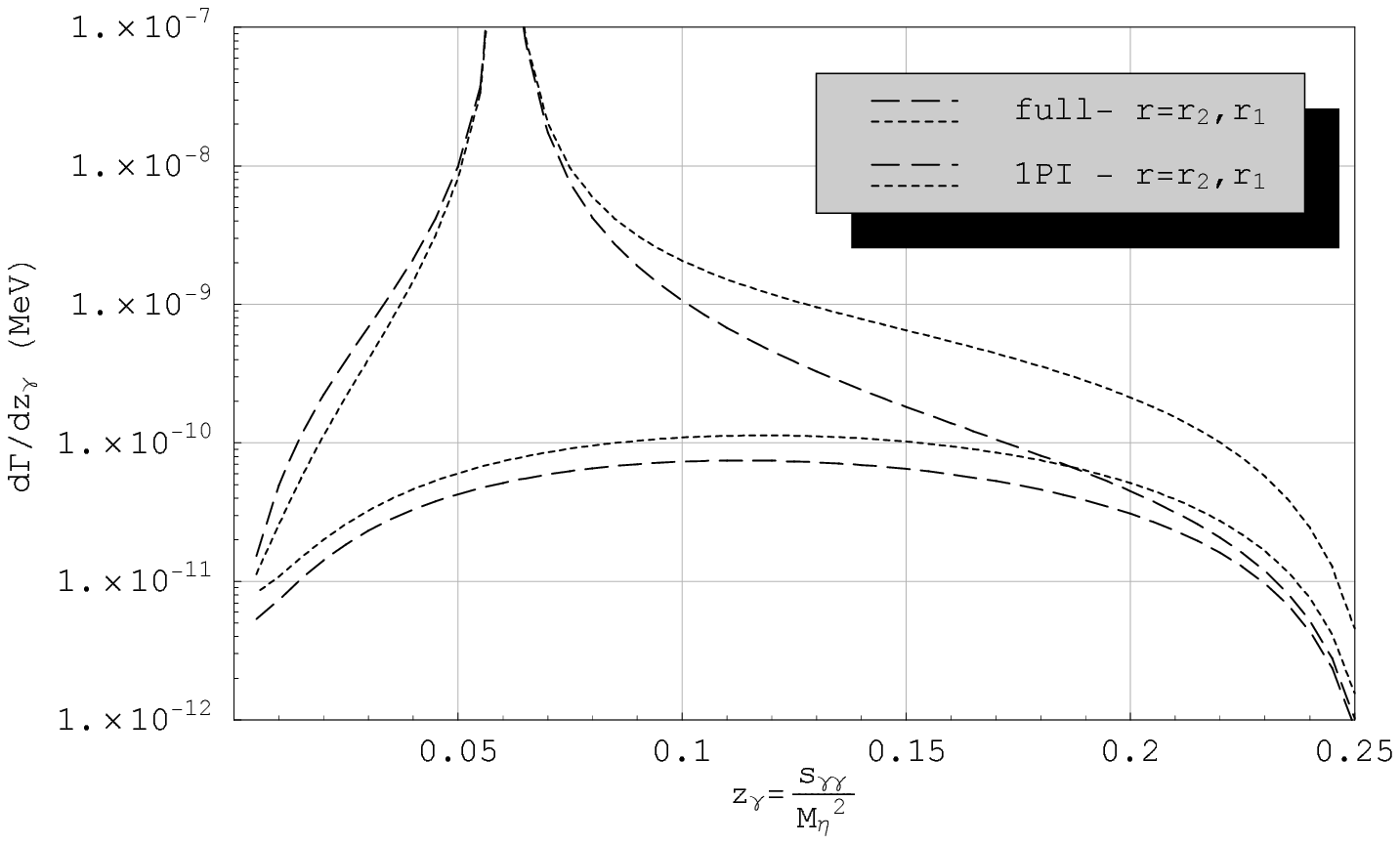,width=0.7\textwidth}
\vspace{-0.25cm}
\caption{One loop corrected full decay width and the 1PI graphs' contribution}
\label{graph3}
\end{figure}
\section{Conclusion}

\no We have analyzed the $\eta\to\pi^0\pi^0\gamma\gamma$ decay to the next-to-leading order of chiral
perturbation theory in its both variants. There are four distinct contributions -- the $\pi^0$-pole, $\eta$-tail, 1PI and counterterms.

The calculation in the Standard limit of the theory proved the dominance of the resonant pion pole contribution in the
kinematic region $s_{\gamma\gamma}<0.20M_{\eta}^2$. The $\eta$-tail graph is small in the whole phase space. The 1PI diagrams can't be omitted, they are dominant for \mbox{$s_{\gamma\gamma}>0.20M_{\eta}^2$}. Our calculations confirm the destructive 1PI-resonant interference in the area $s_{\gamma\gamma}>M_{\pi}^2$.

The Generalized counting brought some important changes. The 1PI contribution could grow around 50\%, the $\eta$-tail up
to 50 times. The latter one could become dominant in the whole region $s_{\gamma\gamma}>0.11M_{\eta}^2$.

In the full partial decay width, the possible violation of the Standard scheme is considerable. Relying upon the work \cite{Ametller}, we neglected the counterterm contribution. We left open the questions about the higher order corrections to the $\eta$-tail amplitude and the experimental value of our calculations.
\\
%  for pictures use \epsfig command
%  for instance:
%\begin{figure}
%\epsfig{figure=myfile.eps,width=0.7\textwidth,height=0.2\textheight}
%\caption{My caption}
%label{My label}
%\end{figure}
\textbf{Acknowledgments}: This work was supported by program
`Research Centers' (project number LN00A006) of the Ministry of
Education of Czech Republic.

\end{document}